%% file: main.tex
\documentclass[conference,9pt]{IEEEtran}
\IEEEoverridecommandlockouts

\usepackage[utf8]{inputenc} % allow utf-8 input
\usepackage{pifont} % for circled numbers
\usepackage{xcolor}
\usepackage{times}
\usepackage{graphicx}
\usepackage{tabularx}
\usepackage{algpseudocode}
\usepackage[compress]{cite}
\usepackage{amsmath,amsfonts,amssymb}
\usepackage{listings}
\usepackage{mathtools}
\usepackage{multirow}
\usepackage{subcaption}
\usepackage{url}
\usepackage{array}
\usepackage{kotex}
\usepackage[linesnumbered,ruled,vlined]{algorithm2e}
\usepackage{diagbox}
\usepackage{tikz}
\usepackage{booktabs}
\usepackage{array}
\newcolumntype{P}[1]{>{\centering\arraybackslash}p{#1}}

\def\figref#1{Fig.~\ref{#1}}

\def\tabref#1{Table~\ref{#1}}

\long\def\ignore#1{}

\def\bottomoffigure{}  %\vspace{-5mm}}
\newcommand{\myincludegraphics}[3][]{\centering
\includegraphics[#1]{#2}\caption{#3\label{#2}}\bottomoffigure}
\graphicspath{{figs/}}

\usepackage[backgroundcolor=green!20]{todonotes}
\usepackage{todonotes}
\newcounter{todocounter}
\makeatletter\@ifpackageloaded{todonotes}{\newcommand{\todonum}[2][]{\stepcounter{todocounter}\todo[inline]{\thetodocounter: #2\ifx\x#1\x{}\else{ [#1]}\fi}}}{\newcommand{\todonum}[2][]{\par\stepcounter{todocounter}\noindent\fcolorbox{black}{green!20}{\begin{minipage}{.98\linewidth}\thetodocounter: #2\ifx\x#1\x{}\else{ [#1]}\fi\end{minipage}}}}\makeatother

\makeatletter\@ifclassloaded{IEEEtran}
{\newenvironment{keyword}{\begin{IEEEkeywords}}{\end{IEEEkeywords}}
\ifCLASSOPTIONconference\renewcommand\IEEEkeywordsname{Keywords}\fi}{
\def\bstctlcite{\@ifnextchar[{\@bstctlcite}{\@bstctlcite[@auxout]}}
\def\@bstctlcite[#1]#2{\@bsphack\@for\@citeb:=#2\do{%
\edef\@citeb{\expandafter\@firstofone\@citeb}%
\if@filesw\immediate\write\csname #1\endcsname{\string\citation{\@citeb}}\fi}%
\@esphack}}\makeatother

\begin{document}

\title{Hyperdimensional Computing as a Rescue for Efficient Privacy-Preserving Machine Learning-as-a-Service%
\thanks{This work was supported by the Samsung Advanced Institute of Technology (SAIT), Samsung Electronics Co., Ltd., by IITP Grant (No.~2020-0-01336, Artificial Intelligence Graduate School Program (UNIST)), IITP Grant, by National R\&D Program through NRF of Korea funded by the Ministry of Science and ICT (RS-2023-00258527), and by the Free Innovative Research Fund of UNIST (1.170067.01).}
\thanks{J.~Lee and H.~Moon are the corresponding authors of this paper.}
}
%\title{Hyperdimensional Computing for Efficient Privacy-Preserving Machine Learning-as-a-Service}

%\title{Hyperdimensional Computing to the Rescue: Efficient Privacy-Preserving Machine Learning-as-a-Service}               

%\title{\huge Ensuring Model Secrecy for On-Device Inference with Homomorphic Encryption and Hyperdimensional Computing}
%{Enabling On-Device Homomorphically Encrypted Inference with Hyperdimensional Computing
%Ensuring Model Secrecy for on-Device Inference with Homomorphic Encryption and Hyperdimensional Computing
%HE-HD: Homomorphically Encrypted- Hyperdimensional Computing for Affordable PPML
%Privacy-Preserving Machine Learning%
%\thanks{ack}

%\author{Jaewoo, Chenghao, Prof. Moon,  Jongeun}
\author{%
  \IEEEauthorblockN{%
    Jaewoo Park\IEEEauthorrefmark{1},
    Chenghao Quan\IEEEauthorrefmark{2},
    Hyungon Moon\IEEEauthorrefmark{1} and
    Jongeun Lee\IEEEauthorrefmark{2}
  }%
 \IEEEauthorblockA{\IEEEauthorrefmark{1}Department of Computer Science and Engineering,
 \IEEEauthorrefmark{2}Department of Electrical Engineering}%
\IEEEauthorblockA{Ulsan National Institute of Science and Technology (UNIST), Ulsan, Korea}%
\IEEEauthorblockA{\texttt{\{hecate64,quanch,hyungon,jlee\}@unist.ac.kr}}%
}

\maketitle

\bstctlcite{IEEEexample:BSTcontrol}

\input{body}

%\clearpage

\bibliographystyle{IEEEtran}
\bibliography{refs}

\end{document}

%% file: body.tex
% 250 words limit
\begin{abstract}
Machine learning models are often provisioned as a cloud-based service where the clients send their data to the service provider to obtain the result. This setting is commonplace due to the high value of the models, but it requires the clients to forfeit the privacy that the query data may contain. Homomorphic encryption (HE) is a promising technique to address this adversity. With HE, the service provider can take encrypted data as a query and run the model without decrypting it. The result remains encrypted, and only the client can decrypt it. All these benefits come at the cost of computational cost because HE turns simple floating-point arithmetic into the computation between long (of degree $\geq$ 1024) polynomials. Previous work has proposed to tailor deep neural networks for efficient computation over encrypted data, but already high computational cost is again amplified by HE, hindering performance improvement. In this paper we show hyperdimensional computing can be a rescue for privacy-preserving machine learning over encrypted data. We find that the advantage of hyperdimensional computing in performance is amplified when working with HE. This observation led us to design HE-HDC, a machine-learning inference system that uses hyperdimensional computing with HE. We carefully structure the machine learning service so that the server will perform only the HE-friendly computation. Moreover, we adapt the computation and HE parameters to expedite computation while preserving accuracy and security. Our experimental result based on real measurements shows that HE-HDC outperforms existing systems by $26\sim3000\times$ times with comparable classification accuracy.
\end{abstract}


%\maketitle

% WHY HDC for PPML?
% competitors of this paper: FHE(CKKS, TFHE) based CNN/DNN
% why not SVM or other neuromorphic algorhitms?
%\todonum{Any suggestions for a better title?}
%\todonum{Ensuring Model Secrecy for on-Device Inference with Homomorphic Encryption and Hyperdimensional Computing}

\section{Introduction}

Machine learning models are frequently deployed as cloud-backed services, wherein clients send their data to the model owner's service to obtain the inference result. This model provision method, known as Machine Learning-as-a-Service (MLaaS), is widespread for several reasons. Clients often prefer not to run the model locally due to limited computational resources or energy constraints. Additionally, model owners prefer to safeguard their models or any associated knowledge, such as training data, from potential leaks.

This prevalent service structure suffers from a critical drawback: the clients' data is vulnerable to exposure by the model owner. Despite employing contracts and security measures, such data is often considered at risk of being leaked. For instance, many tech companies prohibit their employees from using external MLaaS platforms due to concerns over highly confidential assets being compromised. To address these concerns, privacy-preserving machine learning (PPML) techniques~\cite{mp-fhe-cnn,cryptonets:icml16,hesamifard2018privacy} have emerged, with the primary goal of safeguarding the clients' data from being exposed to the model owner.

The PPML techniques for this problem---client data secrecy in MLaaS---can be classified into two groups. The first group~\cite{shadownet} is the TEE (trust execution environment) or enclave approach, where MLaaS runs their model within a secure enclave on their cloud platform so that clients can attest that the enclave does not leak their data.
This approach has performance advantages as it can perform inference over plaintext and leverage accelerators and protection mechanisms to enhance performance.
However, the confidentiality of user data hinges upon the strength of the isolation mechanism or the secrecy of the cryptographic keys upon which the enclaves are built.
While the enclave approach holds promise, no existing system is known to be completely secure against side-channel attacks~\cite{chen2021voltpillager,murdock2020plundervolt,chen2019sgxpectre}.

The second group~\cite{gazelle,she:nips18,cryptonets:icml16,shadownet,hesamifard2018privacy} is the cryptographic approach that utilizes \textit{Homomorphic Encryption} (HE).
The HE-based approach has a strong theoretical foundation that ensures data confidentiality on the fly; the only way to compromise data is by stealing the cryptographic key that is kept by the client locally. However, the primary obstacle preventing the widespread use of HE in various applications is its high computational cost, typically 5-6 orders of magnitude slower compared to plaintext operations. Most of the work is dedicated to the systems research \cite{f1accel,nttpim} while not much work focuses on developing HE-friendly algorithms. Thus in this paper, we present a novel strategy for utilizing HE in the context of PPML for classification applications, thereby addressing this challenge.

This paper presents HE-HDC, an efficient yet privacy-preserving ML framework that significantly expedites online MLaaS with HE and \textit{hyperdimensional computing}. 
%We first find the hyperdimensional computing as a promising way to implement MLaaS over HE. 
Hyperdimensional computing (HDC) \cite{kanerva2009hyperdimensional} is an emerging ML paradigm with very efficient training and robustness to noise, in which a model is represented as a set of \textit{hypervectors}. 
An HDC-based classification system, for instance, would first \textit{encode} input data into a hypervector and compare it with model hypervectors. Then the class label associated with the most similar model hypervector is returned as inference result. 
%further details about HDC. 

Our key observation is that when computed over HE, the performance advantage of HDC over DNN (deep neural network) can be greatly amplified. While the performance of DNN-based PPML is limited by excessive computational cost of HE especially for some essential operations (e.g., multiplication and non-linear operations in the case of CKKS), our HE-HDC framework very scarcely uses those operations that are expensive on HE while providing similar privacy guarantee. Also the application domain of HDC is fast expanding, including not just classification but also object recognition~\cite{luczak2022combining}, natural language processing (NLP)~\cite{liu2022l3e, thapa2021spamhd}, reinforcement learning~\cite{chen2022darl}, and bio-informatics~\cite{burrello2018one, kim2020geniehd, ma2021molehd}.
%This is primarily due to the excessive cost of multiplication over HE.

We implemented HE-HDC using the SEAL~\cite{sealcrypto}, which is one of the most widely used library that implements the core of CKKS scheme. The experiment running HE-HDC end-to-end with MNIST data set exhibits that HE-HDC outperforms the existing mechanisms~\cite{cryptonets:icml16,mp-fhe-cnn} running CKKS-tailored deep neural networks by $26 \sim 3000\times$ with comparable classification accuracy.

This paper makes the following contributions:
\begin{itemize}
%two main ideas..
    \item To the best of our knowledge, we are the first to propose and explore using hyperdimensional computing over HE for privacy-preserving MLaaS. Our study shows that our system, HE-HDC, performs image classification tasks with significantly low latency at comparable accuracy when compared to the state-of-the-art PPML techniques.
    \item We identify the inference using hyperdimensional computing requires nearly no multiplication, which is considerably more expensive when computed over HE. We further adapt the inference procedure to eliminate the multiplication operation, enabling to use the optimal parameters for HE.    
\end{itemize}

\section{Related Work}
Here we briefly review previous work on PPML focusing on image classification task and on HE. However, none of the previous work considers combining HE and HDC, nor does any HDC work propose application of cryptography for PPML.

\subsection{Privacy-Preserving Image Classification}
Bourse et al. \cite{she:nips18} present a framework for the homomorphic evaluation of neural networks named FHE-DiNN, in which each neuron’s output is refreshed through bootstrapping.
Juvekar et al. \cite{gazelle} develop Gazelle, a scalable and low-latency system for secure neural network inference, which contains a homomorphic encryption library, homomorphic linear algebra kernels, and optimized encryption switching protocols.
Dowlin et al. \cite{cryptonets:icml16} present CryptoNets, neural networks that can be applied to encrypted data to make accurate predictions while maintaining data privacy and security.
Chou et al. \cite{chou2018faster} present Faster CryptoNets, which accelerate the homomorphic evaluation by developing a pruning and quantization approach that leverages sparse representations. Additionally, they approximate modern activation functions.
Brutzkus et al. \cite{brutzkus2019low} present Low-Latency CryptoNets (LoLa), which change representations of the data throughout the computation by novel ways to reduce latency while maintaining accuracy and security. They also apply the method of transfer learning to provide private inference services.
Boemer et al. \cite{boemer2019ngraphhe} introduce nGraph-HE, a graph compiler, which enables the deployment of trained models with popular DL frameworks while simply treating HE as another hardware target. Furthermore, they develop HE-aware graph-compiler optimizations, both at compile-time and at run-time.
Hesamifard et al. \cite{ehsan2019} design a privacy-preserving classification for convolutional neural networks over encrypted data, which replaces the commonly used activation functions in CNNs with low-degree polynomials.
Benaissa et al. \cite{benaissa2021tenseal} present TenSEAL, a flexible open-source library, for doing encrypted tensor computation using homomorphic encryption.
Lee et al. \cite{mp-fhe-cnn} reduce the runtime overhead of bootstrapping by a multiplexed packing method, which packs data of multiple channels into one ciphertext in a compact manner. Besides, they propose a faster multiplexed parallel convolution algorithm to reduce the number of required rotations using full ciphertext slots.
Liu et al. \cite{trusted-dnn} propose Trusted-DNN, an overall DNN model protection strategy based on TrustZone and encryption algorithms oriented to the security issues of embedded devices.

\subsection{HDC}
Zou et al. \cite{neuralhd} develop NeuralHD, which is the first HDC algorithm with dynamic and regenerative encoder for adaptive learning. NeuralHD can enhance learning capability and robustness by identifying insignificant dimensions and regenerating those dimensions. Furthermore, they present a scalable learning framework to distribute NeuralHD computation over edge devices.
In \cite{zou2021manihd}, the authors propose ManiHD which provides trainable encoding. It considers non-linear interactions between the features. ManiHD also supports online learning by sampling data and capturing important features in an unsupervised manner.
Imani et al. \cite{voicehd} propose VoiceHD, an efficient and hardware-friendly speech recognition technique using HD computing. It maps preprocessed voice signals in the frequency domain to hypervectors and combines them to compute class hypervectors. Additionally, they extend VoiceHD to VoiceHD+NN which uses a single-layer neural network to improve the resolution of similarity measures.
AdaptHD is proposed in \cite{adapthd}, which is an adaptive retraining method for HD computing. AdaptHD introduces the definition of learning rate in HD computing and proposes a hybrid approach to update the learning rate considering both iteration and data dependency.
%OnlineHD \cite{onlinehd} identifies common patterns in each class hypervector and eliminates model saturation. 
Instead of na\"ive data accumulation during training, OnlineHD \cite{onlinehd} updates the model differently depending on the model prediction result, which enables iterative training, potentially boosting performance of HDC models.
Kim et al. \cite{cascadehd} propose CascadeHD, an efficient many-class classification learning framework, which considers the inter-dependency of different classes by revising the iterative hypervector fine-tuning procedure and identifies confusing classes while learning a hierarchical inference structure by a meta-learning algorithm.
Prive-HD \cite{privhd} may be most similar to our work in that it considers HD computing from a privacy perspective. The authors observe that HD computing has no privacy due to its reversible computation, so they present Prive-HD to realize a differentially private HD model as well as to blur the information sent for cloud-hosted inference by quantization and pruning. However, in our work, the client data is encrypted, so there is no privacy issue.

\section{Preliminary}

\subsection{Hyperdimensional Computing}

The hyperdimensional computing (HDC) model is a machine learning paradigm that is built around vectors of large dimensions ($D \simeq 4000$ or more) called \emph{hypervectors}.  The HDC approach has important advantages including much simpler training (not based on back-propagation), superior generalization performance, and holographic representation promoting noise resilience \cite{rahimi2016hyperdimensional, 7838428, thomas2021theoretical, neuralhd}.  Here we briefly explain HDC using a simple image classification system example, which consists of an encoder and a classifier.

\noindent\textbf{Encoding:} 
The goal of encoding is to find a well-defined hypervector representation for input data. Though various encoding schemes have been proposed in the literature \cite{neuralhd, zou2021manihd, voicehd, adapthd, onlinehd, cascadehd}, they all share the common principle: (i) the generated hypervectors should be nearly orthogonal to one another, and (ii) different data points should result in different hypervectors.
%In practice, the hypervectors has a dimenstion $D\simeq 4000$, large enough to populate orthogonal hypervectors as much as needed.\todo{revise: populate orthogonal hypervectors as much as needed}
%
As an example, random projection encoding~\cite{onlinehd} transforms a flattened image vector $\vec{X}$ of length $K$ into a hypervector $\vec{H}$ of length $D$ as follows.
\begin{equation}
    \vec{H} = \cos(B \vec{X} + \vec{b}) \otimes \sin(B \vec{X})
\end{equation}
where $B$ is a $(D \times K)$-sized matrix, whose columns are random hence orthogonal \emph{base hypervectors} drawn from a standard normal distribution; $\vec{b}$ is another random base hypervector of length $D$ drawn from a uniform distribution in $[0, 2 \pi]$; and $\otimes$ is element-wise multiplication.

%The procedure of hyperdimensional classification is composed of four steps. The first step is encoding, where the given data points are mapped into a high-dimensional space. Secondly, the encoded hypervectors are used in the training step to generate a class hypervector representing each class. During the inference stage the test data is encoded as a hypervector then similarity search between the encoded hypervector and the class hypervectors is performed. The hypervector with the highest similarity is returned as the classification result. Lastly, a retraining step is applied in order to boost the classification accuracy. In the following, each steps are explained in details.

\noindent\textbf{Model and Training:}  
A typical HDC-based image classifier model consists of a number of \textit{class hypervectors} $\vec{\mathcal{C}^l}$, one for each class (or label) $l$.  A class hypervector is a generalized representation of each class.  Due to the orthogonality of hypervectors, training, or computing class hypervectors, is extremely easy, and does not require error back-propagation.  Once input hypervectors are generated, a class hypervector is obtained by accumulating all hypervectors belonging to the class.
\begin{equation}
    \vec{\mathcal{C}^l} = \sum_{j} \vec{\mathcal{H}_j^l}
\end{equation}

\noindent\textbf{Classifier and Inference:}
Classification of a query data is performed by first mapping the query data to a hypervector $\vec{\mathcal{H}}_q$, and then computing the similarity of the \emph{query hypervector} with every class hypervector $\vec{\mathcal{C}^l}$. The label with the highest similarity is selected, which is the inference result.
For the similarity measure ($\delta$), cosine similarity is typically used:
\begin{equation}
    \delta(\vec{\mathcal{H}}_q,\vec{\mathcal{C}^l}) = \frac{\vec{\mathcal{H}}_q^\intercal \cdot \vec{\mathcal{C}^l}}{\Vert \vec{\mathcal{H}}_q \Vert \cdot \Vert \vec{\mathcal{C}^l} \Vert }
\label{cosinesim}
\end{equation}
where $\Vert \cdot \Vert$ represents the $L_2$ norm.

\noindent\textbf{Iterative Training:}
Due to the limited performance (i.e., inference accuracy) of single-pass training, an iterative training method can be employed \cite{onlinehd}, whose idea is to examine mispredicted queries and update relevant class hypervectors (i.e., of correct and mispredicted labels).  Given a query hypervector $\vec{\mathcal{H}_q}$ with the correct label $l$ and the mispredicted label $m$, the class hypervectors are updated as follows ($\eta$ is learning rate):
\begin{align}
\vec{\mathcal{C}^l} &= \vec{\mathcal{C}^l} + \eta (1-\delta(\vec{\mathcal{H}}_q, \, \vec{\mathcal{C}^l})) \vec{\mathcal{H}}_q
\\
\vec{\mathcal{C}^m} &= \vec{\mathcal{C}^m} - \eta (1-\delta(\vec{\mathcal{H}}_q, \, \vec{\mathcal{C}^m}))  \vec{\mathcal{H}}_q
\end{align}

\begin{table*}[t]
  \centering
  \caption{Key terms and symbols used for CKKS}
  \label{terms}
  \begin{tabularx}{\linewidth}{c l}
    \toprule
    \multicolumn{1}{c}{Term} & \multicolumn{1}{c}{Meaning} \\
    \midrule
    Multiplicative Depth & The number of multiplications between ciphertexts that can be performed after one encryption.  \\
    \cmidrule{2-2}
    \multirow{2}{*}{Modulus Chain ($Q = (Q_1, \dots, Q_l)$)} & A sequence of moduli that a CKKS scheme uses for computation with ciphertexts. \\
    & The length of chain determines the multiplicative depth and the sum the level of security. \\
    \cmidrule{2-2}
    \multirow{2}{*}{$N$} & The degree of ciphertext (which is represented as a pair of polynomials). $N$ must be a power of an integer \\
    & and a multiple of two. This parameter affects the level of security and latency of computation with ciphertexts.\\
    %Rotation & Computes\\
    %Bootstrapping & \\
    \bottomrule
  \end{tabularx}
\end{table*}

\subsection{CKKS: an HE Scheme for Approximate Arithmetic}

%\todonum[HM]{Will also mention that multiplication is expensive, with numbers (from reference)}

\begin{figure}
  \myincludegraphics[width=\linewidth]{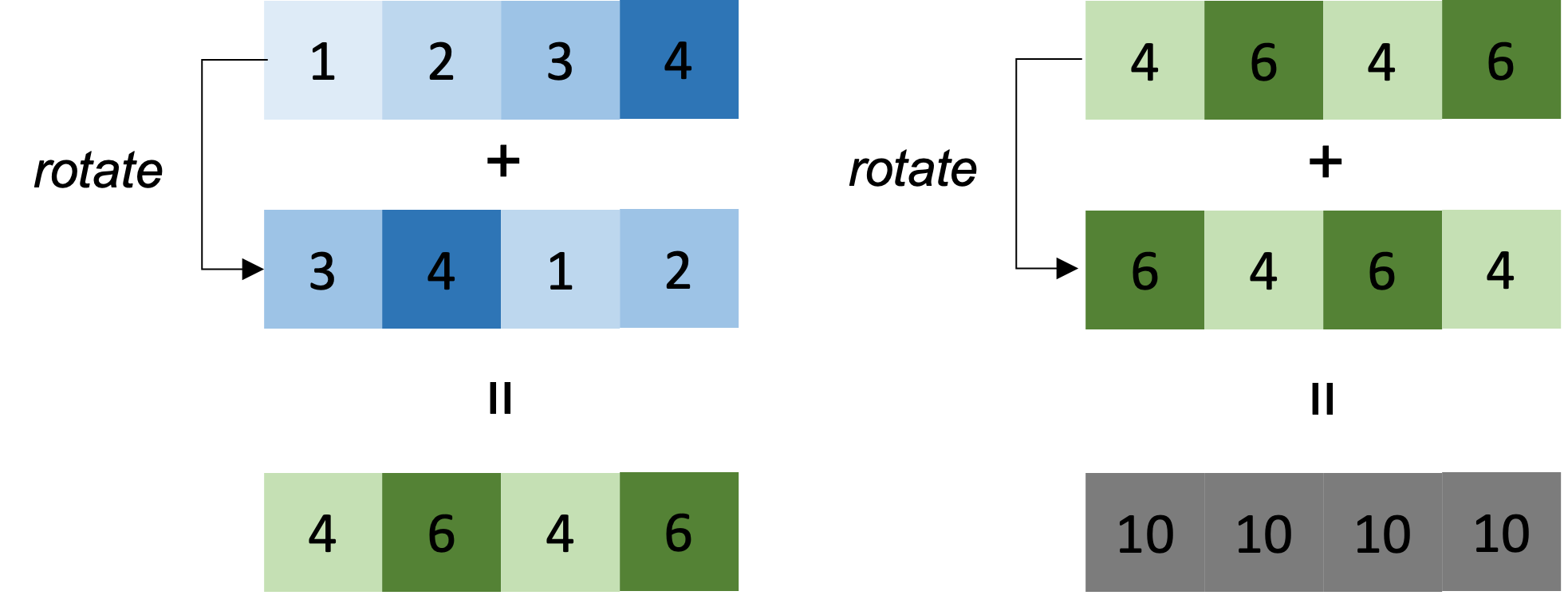}{Homomorphically summing all elements of a plaintext vector (of length-$n$) using $\log_2{n}$ rotations and $\log_2{n}$ additions.}
\end{figure}

CKKS~\cite{ckks:asiacrypt17} is an HE scheme that is tailored for approximate arithmetic such as fixed-point operations. 
A ciphertext, represented as a pair of polynomials $(c_0,c_1)$, each of which is of degree $N$, corresponds to a plaintext of a length-$N/2$ vector of integers. 
To represent fixed-point values as integers, scale factor ($\Delta$) is used, which causes a problem with multiplication (note that both plaintext and ciphertext are represented as integer-coefficient polynomials), requiring a \textit{rescaling operation} after each multiplication.

\ignore{As an HE scheme, it supports addition and multiplication between two encrypted fixed-point numbers. Result of adding or multiplying two encrypted approximate numbers followed by the decryption is the same as adding or multiplying the two numbers without encryption, as shown below.
\begin{equation}
\begin{aligned}
    a + b &= \text{Decrypt}(\text{Encrypt}(a) + \text{Encrypt}(b)) \\
    a \times b &= \text{Decrypt}(\text{Encrypt}(a) \times \text{Encrypt}(b))
\end{aligned}
\end{equation}
One ciphertext in the CKKS scheme is the result of encrypting a vector of real values, and is a tuple of two polynomials $(c_0,c_1)$. The degree of each polynomial is $N$, which has to be a multiple of 2. Such as ciphertext corresponds to a real-valued vector of dimension $N/2$ before encryption or after decryption. }

Other operations provided by CKKS include \textit{bootstrapping} and \textit{rotation}.
Ciphertexts have \textit{noise}, limiting the number of operations they can undergo before correct decryption is impossible. Multiplication increases noise significantly, hence \textit{multiplicative depth}, or the number of homomorphic multiplications a ciphertext can go through, is an important parameter. The operation that effectively reduces this noise is called \textit{bootstrapping}, which is however very expensive. 
To increase multiplicative depth (which is the length of modulus chain minus one), a large ciphertext modulus is required. 

The homomorphic \textit{rotation} operation effectively rotates the corresponding plaintext vector. In practice, we often need to rotate a vector so that we can perform an operation involving an element of a vector and another element of the same or a different vector. \figref{ckks_reducesum} illustrates how to perform a reduce-sum operation on a plaintext (here, a length-4 vector) using a series of homomorphic rotations and additions. 

%\textbf{Bootstrapping and Rotation}
%CKKS provides additional operations that help the efficiency and expressiveness of computation with ciphertext. First is \textit{bootstrapping} that enables to perform multiplication operations beyond the multiplicative depth. Multiplication between ciphertexts in CKKS increases the \emph{noise} that CKKS adds during encryption, and each multiplication operation increases this noise. When the number of multiplication operation exceeds the multiplicative depth, the result becomes impossible to decrypt correctly. Bootstrapping is the operation that effectively reduces this noise so that the result of further multiplication can be decrypted correctly. 
%The Second is the \textit{rotation} that effectively rotates the plaintext vector in a ciphertext. Recall that one ciphertext is the result of encrypting a vector of plaintext. In practice, we often need to rotate the vector so that we can compute using $i$-th element of a vector with $j$-th element of another, where $i \neq j$. Rotation is the primitive operation that the CKKS provides and HE-HDC uses it during similarity computation.

%\textbf{CKKS Parameters}
CKKS is configurable with two parameters ($N$, $Q$) that affect the level of security and computation performance (see \tabref{terms}). These parameters must be chosen so that a certain level of security can be cryptographically provided.  \tabref{parameters} lists the set of parameter combinations we consider in this paper.
Increasing $N$ results in slower computation over encrypted data due to longer ciphertexts. However, decreasing $N$ mandates the sum of $\log_2{Q_i}$ to be reduced in order to maintain the same level of security. Also, decreasing the sum of $Q$ reduces the precision of computation, negatively impacting the quality of computation (e.g., inference accuracy). Throughput this paper we use $\log q$ to denote $\sum_i \log_2 Q_i$.

\begin{table}[t]
  \centering
  \caption{Parameter Combinations Used in Experiments}
  \label{parameters}
  \begin{tabular}{cccc}
    \toprule
    Security Level (bits) & $\log_2 N$ & $\log_2 q$ & $(\log_2 Q_i)$\\
    \midrule
    128 & 11 & 54 & (27, 27)\\
    128 & 12 & 109& (54, 54) \\
    128 & 13 & 218& (60, 60) \\
    128 & 14 & 438& (60, 60) \\
    \bottomrule
  \end{tabular}
\end{table}

\subsection{Threat Model}
%\cite{trusted-dnn,shadownet}

We assume a common threat model that privacy-preserving MLaaS considers~\cite{she:nips18,cryptonets:icml16}. The MLaaS provider offers an image classification service and prefers not to reveal the model itself. For example, the provider aims to keep the class hypervectors as secrets. Clients want to use the classification service using private input data. They do not trust the MLaaS provider, i.e., they assume that the MLaaS provider may want to collect the private data if clients send them to the provider. We also follow a common assumption of the existing HE-based schemes by assuming that the MLaaS provider is honest but curious, i.e., it wants to steal clients' data but performs the requested computation honestly. This is a reasonable assumption in that dishonest computation in a service will simply reduce the quality of service, making it less attractive to clients.

%\todonum{a fancy diagram illustrating threat model}
% HM: Inessential

\section{HE-HDC}

\begin{figure*}
    \myincludegraphics[width=\linewidth]{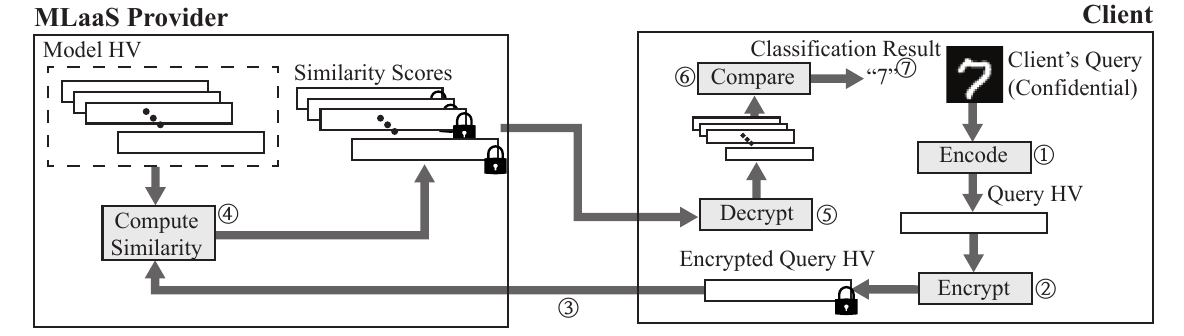}{An overview of HE-HDC. HV is short for hypervector.}
    %HM: Seem we cannot afford further updates.
    
\end{figure*}

%각 시나리오 별로 설명.
\subsection{Overview}

%The encrypted query hypervector is then sent to the server along with a public key and a rotation key.
\figref{he-hdc} gives an overview of our HE-HDC approach. Similar to previous FHE-based PPML approaches, we assume that a fully trained hyperdimensional classifier model runs on a server, i.e., the server has unencrypted class hypervectors as their owner. The encoder of the HDC classifier is shared with the client device, which uses it to \emph{encode} \ding{\numexpr172} the user's (private) data into a query hypervector. The query hypervector is then \emph{encrypted} \ding{\numexpr173} into a ciphertext using the user's \emph{secret key}. The encrypted query hypervector is then sent \ding{\numexpr174} to the server. The MLaaS provider homomorphically computes similarity scores \ding{\numexpr175} for all classes, resulting in $L$ encrypted similarity scores \ding{\numexpr176} where $L$ is the number of classes. These encrypted scores are sent back to the client, which decrypts \ding{\numexpr177} the scores by using the secret key. Note that the result of any homomorphic computation using encrypted data (e.g., the query hypervector) remains encrypted and only the client, who has the secret key, can decrypt it. As the last step, the client compares the similarity scores \ding{\numexpr178} to obtain the class result \ding{\numexpr179}.

%\subsection{Security Guarantee}

\noindent\textbf{Rationale for Client-side Encoding: }
Exposing the encoder that consists of the base hypervectors does not enable the clients to learn about the provisioned model. These hypervectors are generated randomly in principle~\cite{kanerva2009hyperdimensional}, and the choice of them does not affect the accuracy of the trained classification model significantly.\footnote{We have generated 100 different encoders and used them to train HDC classifiers for MNIST, which have shown very uniform classification accuracy with variance of less than $10^{-3}$\%.} For this reason, revealing the base hypervectors does not help the clients to infer the class hypervectors, which is the trained model.
%Our HDC classifier is the same on proposed by \cite{onlinehd}.

%\subsection{Opportunities and Challenges in HDC over CKKS}

%\textbf{Opportunity}
%Inference in HDC is a good fit to run over CKKS in that the computation is composed only of arithmetic operations of approximate numbers. Unlike deep neural networks that have non-linear layers (e.g., ReLU), we do not need to introduce any further approximation to run HDC over CKKS. Avoiding this, performing HDC over CKKS does not affect the inference accuracy. Moreover, computing cosine similarity, the only operation that the MLaaS provider must perform over encrypted data using the client's query, requires only one level of multiplication. As mentioned earlier, multiplication by itself is expensive over CKKS, and having many multiplications require the parameter choices with high multiplicative depth, slowing down the entire encrypt-compute-decrypt flow.

%\subsection{Computing Cosine Similarity over CKKS}

% HM: What is left to the server and its complexity – why is it non-trivial to compute cosien similarity over CKKS?
\noindent\textbf{Computing Cosine Similarity over CKKS: }
The task left for the MLaaS provider, or the server, is to compute the cosine similarity between the query hypervector and each class hypervector, as~\eqref{cosinesim} shows.
Each cosine similarity computation requires one dot product between two hypervectors and two scalar divisions. Assuming the number of slots inside a single ciphertext is no less than the dimension of each hypervector, the dot product could be implemented as a single ciphertext-plaintext multiplication followed by $\log_2{D}$ ciphertext rotations and $\log_2{D}$ ciphertext-plaintext additions. Division by $\Vert \vec{\mathcal{C}^l} \Vert$ requires additional ciphertext-plaintext division. However the division by $\Vert \vec{\mathcal{H}}_q \Vert$ requires ciphertext-ciphertext division, which is not supported by CKKS, as well as many ciphertext-ciphertext multiplications and additions to compute L2-norm.
%(Just N + 1 mults and N - 1 adds and 1 div?)
%Or to be more exact:
% N+1 ciphertext-plaintext mult (N for numerator, 1 for denominator)
% N-1 ciphertext-ciphertext add
% ignoring plaintext operation for computing $\Vert \vec{\mathcal{C}^l} \Vert$
% 1 ciphertext-ciphertext div
% N ciphertext-ciphertext mult, N-1 ciphertext-ciphertext add (for $\Vert \vec{\mathcal{H}}_q \Vert$)

\subsection{Optimizing Similarity Search on CKKS}
\label{optimization}
%In order to evaluate the difference between a query hypervector and class hypervectors, the cosine similarity is used in HDC. The cosine similarity is defined as \eqref{cosinesim}, a dot product between two vectors divided by the $L_2$ norm of the two vectors.

%\subsubsection{Eliminating Homomorphic Division}
\noindent\textbf{Eliminating Homomorphic Division:}
Unlike homomorphic multiplication, homomorphic division is not supported natively by CKKS. To avoid division in similarity search, we perform these two optimizations.

First, we store class hypervectors in a normalized form $\vec{\mathcal{C}'^l} = \vec{\mathcal{C}^l}/\Vert \vec{\mathcal{C}^l} \Vert$, which can save one homomorphic multiplication as shown below.
\begin{equation}
\label{normsim}
    \delta(\vec{\mathcal{H}}_q,\vec{\mathcal{C}^l}) = \frac{\vec{\mathcal{H}}_q^\intercal \cdot \vec{\mathcal{C}^l}}{\Vert \vec{\mathcal{H}}_q \Vert \cdot \Vert \vec{\mathcal{C}^l} \Vert } = \frac{\vec{\mathcal{H}}_q^\intercal \cdot \vec{\mathcal{C}'^l}}{\Vert \vec{\mathcal{H}}_q \Vert}
\end{equation}

Second, we observe that the computation of $\Vert \vec{\mathcal{H}}_q \Vert$ is not strictly necessary for the purpose of classification. We only need to do comparison among labels, and $\Vert \vec{\mathcal{H}}_q \Vert$, being common for all labels, does not affect the comparison result, and therefore can be eliminated. 
After these two simplifications, similarity search is reduced to just one dot-product operation per class.

%\subsubsection{Matrix-Vector Multiplication Optimization}
\noindent\textbf{Matrix-Vector Multiplication Optimization:}
%\todonum[JL]{does using a matrix form have actual performance advantage?  If so, it needs to be mentioned here as well}
To expedite the computation of cosine similarity, we stack class hypervectors to form a class hypervector matrix $M$ and perform a single matrix-vector multiplication (MVM) between the query hypervector and the class hypervector matrix as shown below.
\begin{align}
M &= \begin{bmatrix}
\vec{\mathcal{C}'^1} &
\vec{\mathcal{C}'^2} &
\cdots &
\vec{\mathcal{C}'^L}
\end{bmatrix} \\
    \vec{\delta}(\vec{\mathcal{H}_q}, M) &= \vec{\mathcal{H}}_q^\intercal \cdot M
\end{align}
This allows us to leverage the highly optimized MVM library within CKKS. Given that MVM is one of the most widely used forms of computation, it has gained significant attention from the community for optimization. In particular, we employ the dense vector-row major matrix multiplication method proposed as proposed by Brutzkus et al.~\cite{brutzkus2019low}.

%Most of the CKKS-based PPML works try to optimize latency by finding a more suitable mapping of tensors to ciphertexts \cite{mp-fhe-cnn, brutzkus2019low}.
% JW: wanted to mention that the way of packing matrices into ciphertexts have an impact on performance
%However, in the case of HDC, the class hypervector matrix is an extreme oblong, the column dimension being orders of magnitude greater than the row dimension. The multiplication with such a matrix can hardly benefit from optimized MVM libraries.

%\todonum[HM]{Should blend these three paragraphs into the above.}

%\noindent\textbf{Encrypting a Hypervector:}
One hypervector is encrypted into one or more ciphertexts depending on the hypervector dimension and the number of slots in a single ciphertext ($N/2$).

\noindent\textbf{Case 1:}
When the number of slots inside a single ciphertext ($N/2$) is no less than the hypervector dimension ($D$), each hypervector can be encrypted into a single ciphertext (or a plaintext in the case of $\vec{\mathcal{C}'^l}$).
The dot-product between plaintext and ciphertext is evaluated by a ciphertext-plaintext multiplication followed by a reduce sum of the resulting ciphertexts as illustrated in \figref{ckks_reducesum}.

\noindent\textbf{Case 2:}
Otherwise, $k = \lceil \frac{2D}{N} \rceil$ ciphertexts must be used for a hypervector.
%When the available slots inside a ciphertext is less than the hypervector dimension, a hypervector is encrypted as $\lceil \frac{2D}{N} \rceil$ ciphertexts.
We do element-wise multiplication of two hypervectors by first performing ciphertext-wise multiplications, generating $k$ ciphertexts, which are added together to generate one ciphertext, to which a reduce sum is applied to obtain the final result. Note that doing ciphertext additions first then a reduce sum is significantly faster than the opposite case (i.e., reduce sums for all ciphertexts followed by ciphertext additions), since it involves fewer ciphertext rotations. 

Our inference latency result for the MNIST dataset is $57\times$ faster than the DNN-based PPML approaches such as \cite{brutzkus2019low}, even though we have not employed the state-of-the-art MVM algorithm (e.g., \cite{9665288}). 

%\subsection{Comparison of Several MVM Mappings on CKKS} 
%\subsection{Naive MVM implementation}
%\begin{figure}
%\label{naive_search}
%\myincludegraphics[width=\linewidth]{naive_search}{The naive algorithm to compute cosine similarity search with 3 class hypervectors.}
%\end{figure}

\begin{figure}
  \myincludegraphics[width=\linewidth]{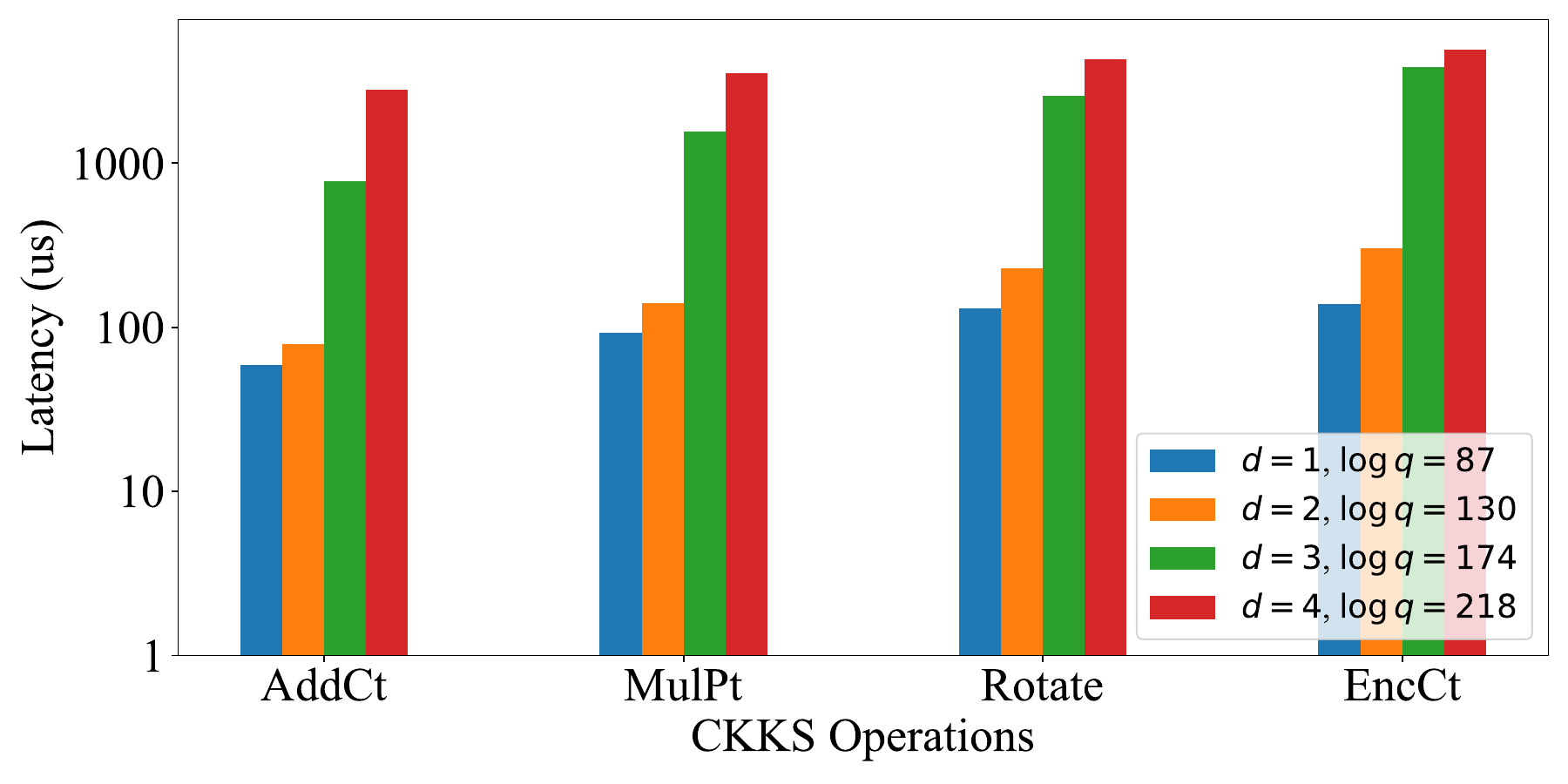}{Latency of CKKS operations for various multiplicative depths ($d$) when we choose $N$ as 8192 and the largest $Q$ that satisfies the 128-bit security.}
  %\todonum[HM]{@JW: please check if the red parts are correct.}

\end{figure}

\subsection{Optimization of CKKS Parameters}
%Choice of the Modulus Chain}

%\todonum[JL]{move to elsewhere}
%In the same way, the rescaling operation could also be ignored after the ciphertext-plaintext multiplication which benefits the performance. 

As shown in \figref{depth_latency}, the latency of CKKS operations varies significantly depending on the length of modulus chain ($Q$), which is determined by the multiplicative depth $d$ required by an application.
While DNN-based PPML approaches require a multiplicative depth of up to 14 \cite{mp-fhe-cnn}, our HE-HDC requires only a single ciphertext-plaintext multiplication in any data flow path. Moreover, this multiplication need not be followed by a rescaling operation, because the multiplication increases the scale factor of \textit{all ciphertexts universally}. Therefore there is no need for rescaling if absence of overflow can be guaranteed, which is the case in our application.
With this \textit{rescale skipping} optimization, we can use the shortest modulus chain possible, with just two modulus integers. For a given polynomial degree ($N$), the runtime of CKKS operations is not affected by modulus bit-widths. Thus we choose the largest modulus chain bit-widths ($\log q$) that meet the 128-bit security standard. 

\begin{figure}[t]
%\label{ckkslatency}
  \myincludegraphics[width=\linewidth]{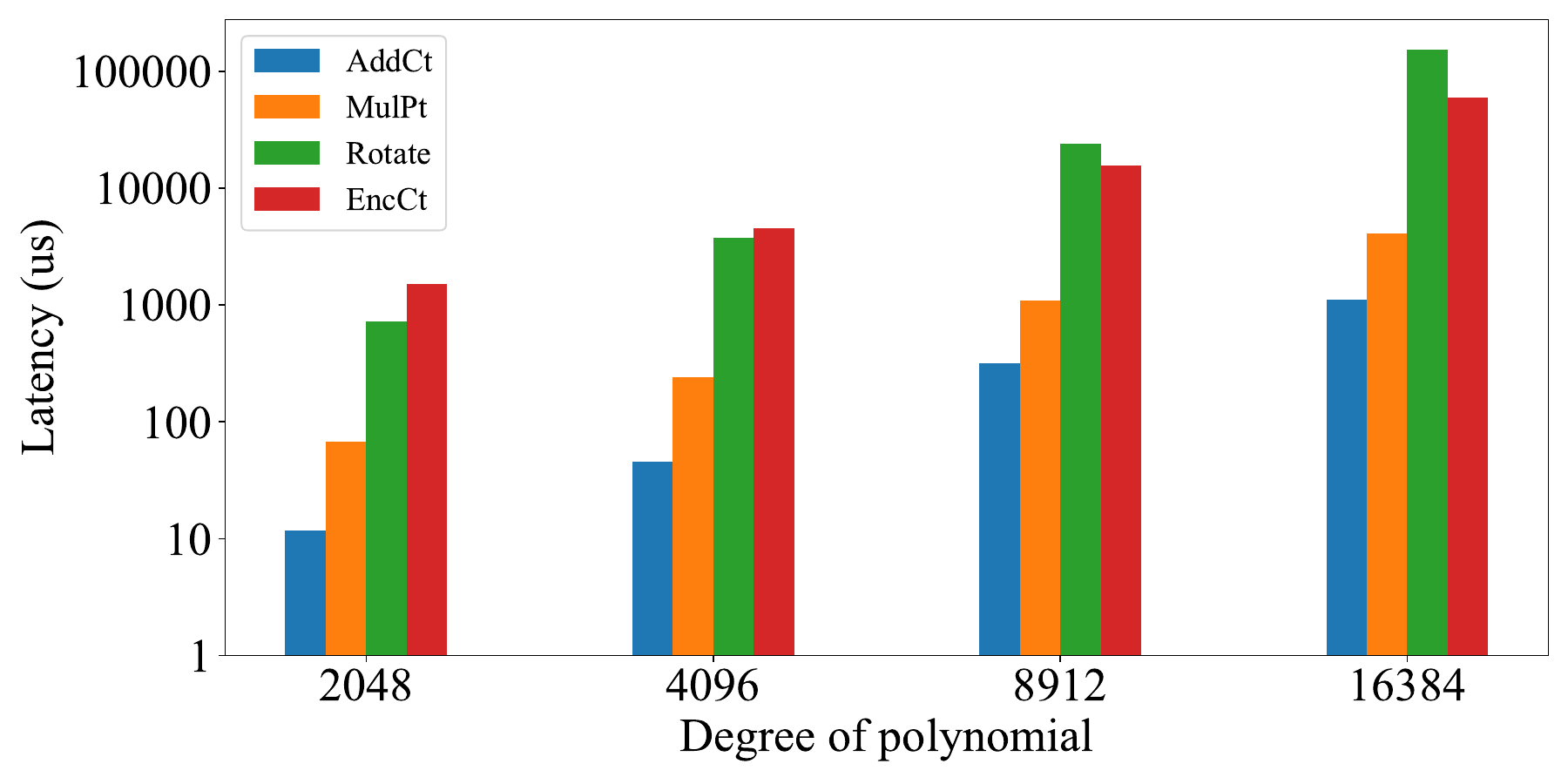}{The latency of CKKS operations with respect to the polynomial degree $N$.}
\end{figure}

%\subsection{Effect of Quantization}
\subsection{Keeping the Polynomial Degree Low with Quantization}

\figref{degree_ops_latency} shows that the latency of CKKS operations grows exponentially with respect to the degree of polynomials. To achieve better performance, it is desirable to use short polynomial degrees. However, using low-degree polynomials prevents us from choosing a large value for $\log q$, which affects the precision of computation with CKKS. For example, the $\log q$ must not be larger than 54 when the degree of polynomial ($N$) is 4096. The corresponding modulus chain $Q$ with the highest arithmetic precision is (27, 27). Under a modulo integer of only 27 bits, the HDC classification accuracy is significantly low. To achieve both high model accuracy and high performance under limited arithmetic precision, we quantize the class hypervectors. \figref{modulo_quant} illustrates the effect of quantization on model accuracy for different modulo bit-widths for a hypervector dimension $D=2048$. The encryption of floating-point vectors into CKKS ciphertexts already acts as an fixed-point quantization, multiplying the floating-point values by $2^{\Delta}$ and rounding to the nearest integer. In HE-HDC, the CKKS scale is set to be greater than the quantization bit-precision of query hypervectors. Using an exhaustive search we determine the scale factor to have the best model accuracy. 

%Despite the limited precision of lower polynomial lengths, \figref{modulo_quant} shows 
\begin{figure}
  \myincludegraphics[width=\linewidth]{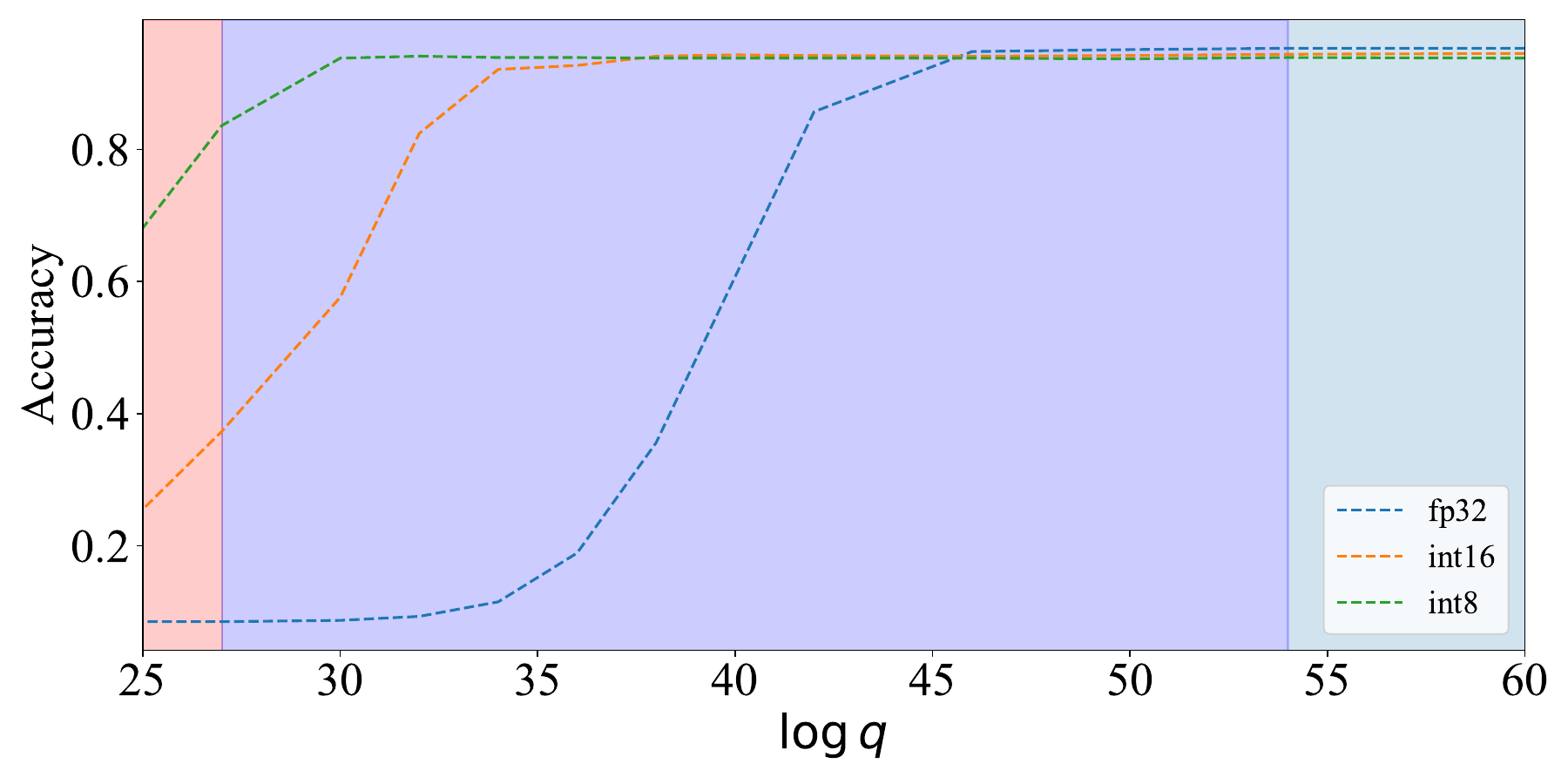}{Effect of quantization in accuracy with respect to modulus bit-length. The three colored regions indicate polynomial degrees ($N = 2^{11}, 2^{12}, 2^{13}$).}
\end{figure}

%\begin{figure}
%\label{ms_ckks_parameters}
%  \myincludegraphics[width=\linewidth]{ms_ckks_parameters}{The maximum secure $Q_i$ for polynomial length and multiplication depth}
%\todonum[]{I copied this figure from the link below. We need to redraw this using more lines for multiplication depth = 1, 2, 3...}
%https://www.microsoft.com/en-us/research/publication/protecting-privacy-through-homomorphic-encryption/
%\end{figure}

\section{Experiments}
\subsection{Experimental Setup}

%\subsection{Sensitivity to Random Matrix}
We have implemented our HE-HDC using the Microsoft SEAL library \cite{sealcrypto} to evaluate the inference latency and classification accuracy. \tabref{parameters} shows the CKKS parameters used in our experiments. As for the other configuration or parameters, we use the default values of the SEAL library. A single Intel i5-1038NG7 CPU without multithreading is used to measure the latency of both inference and encryption (on the client). The training and quantization of HE-HDC is implemented extending the official code of Online-HD \cite{onlinehd}.

\begin{table}[t!]
\centering
%  \scriptsize
  \caption{Impact of Design Decisions}
  \label{designdecision}
  \begin{tabular}{ l | l }
  \toprule
  \multicolumn{1}{c|}{Method}  & Inference Latency (ms) \\
  \midrule
  Without Normalization and with Rescale$^\dagger$  & ~~~257.37 ($\times$1.30)\\
  Without Normalization (Rescale Skipped) & ~~~214.75 ($\times$1.09) \\
  With Normalization (Rescale Skipped) &  ~~~197.41 ($\times$1) \\
  \bottomrule
\multicolumn{2}{l}{\emph{Note.}~ 1. $N=8192$ and $D=4096$.} \\
\multicolumn{2}{l}{~~~~~~~  2. $^\dagger$Requires a longer modulus chain.}
  \end{tabular}
\end{table}

\begin{figure*}
\myincludegraphics[width=\linewidth]{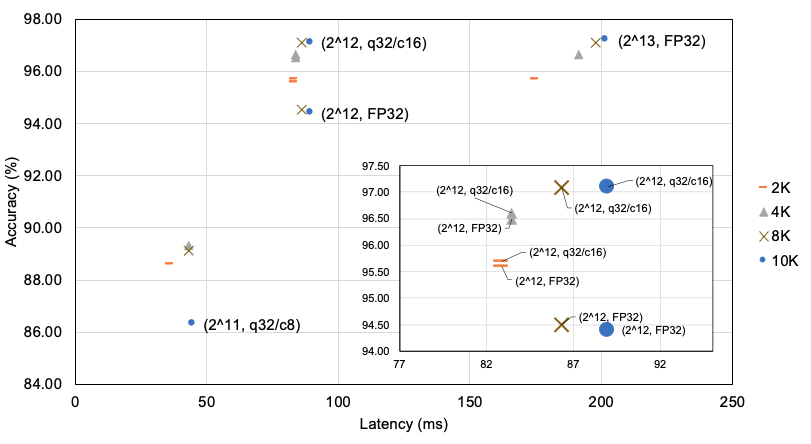}{
The impact of quantization on classification accuracy when using various polynomial degrees ($N = 2^{11}, 2^{12}, 2^{13}$) and HV dimensions ($D$ = 2K, 4K, 8K, 10K, depending on the marker symbol). Inset shows the results of $N=2^{12}$. q32/c16 means 32-bit query HV and 16-bit class HVs.}
\end{figure*}

%2^13 = 8192
\subsection{Inference Latency}
\tabref{designdecision} summarizes the effect of our optimizations (class hypervector normalization and rescaling skipping). With the polynomial degree of $2^{13}$, the inference latency of a na\"ive implementation takes around 257 ms. By removing the \textit{rescaling} operation and therefore shortening the modulus chain, latency decreases by 21\%. Further normalizing the class hypervectors improves the latency by an additional 9\%.

The inference latency for different hypervector dimensions ($D$) and polynomial degrees ($N$) of ciphertexts is presented in \figref{degree_dim_sensitivity}. The latency is more sensitive to the polynomial degree than the hypervector dimension because the rotation is the most time-consuming operation whose performance is only affected by the polynomial degree. We observe that it is always better to lower the  polynomial degree as long as the arithmetic precision is sufficient.
\begin{figure}
  \myincludegraphics[width=\linewidth]{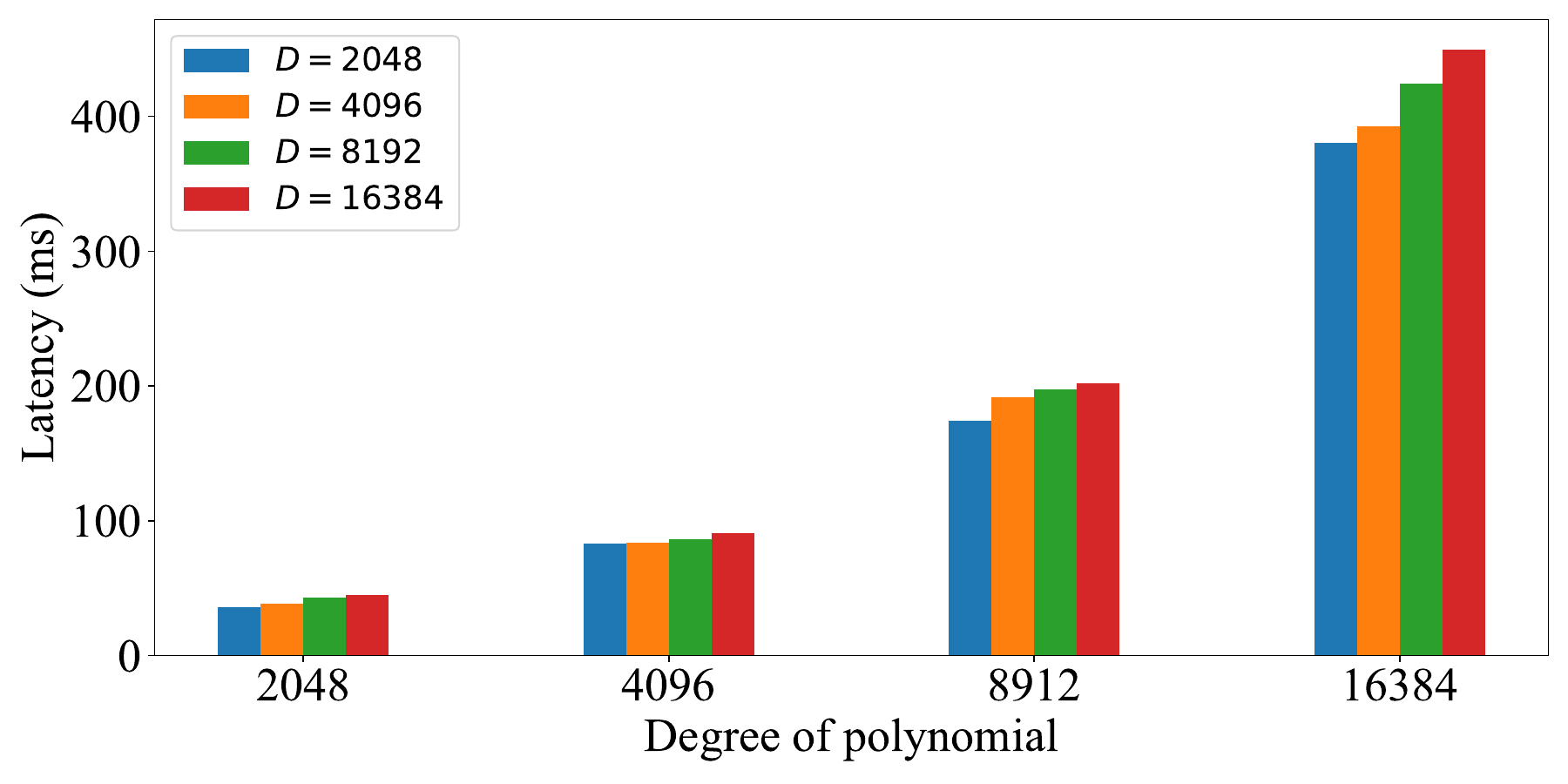}{Inference latency with various polynomial degrees ($N$) and hypervector dimensions ($D$).}
\end{figure}

\ignore{
\subsection{Classification Accuracy}
% \usepackage{multirow}
\begin{table*}[ht]
\begin{tabular}{ccccccccccc}
\hline
  &                       &                             & \multicolumn{8}{c}{D}                                                                                                            \\
  &                       &                             & \multicolumn{2}{c}{2048}           & \multicolumn{2}{c}{4096}          & \multicolumn{2}{c}{8192} & \multicolumn{2}{c}{10240}    \\ \hline
  &                       & \multicolumn{1}{c|}{}       & Acc. (\%) & Latency (ms)           & Acc. (\% & Latency (ms)           & Acc. (\%  & Latency (ms) & Acc. (\% & Latency (ms)      \\
N & 8192                  & \multicolumn{1}{c|}{fp32}   & 95.7      & 174.39                 &          & 191.57                 & 97.1      & 197.65       &          &                   \\
  & 4096                  & \multicolumn{1}{c|}{fp32}   & 95.7      & 82.88                  &          & 93.45                  & 97.1      & 91.00        &          &                   \\
  & \multirow{2}{*}{2048} & \multicolumn{1}{c|}{a32/b8} & 95.4      & \multirow{2}{*}{35.60} &          & \multirow{2}{*}{38.29} &           & 42.78        &          & \multirow{2}{*}{} \\
  &                       & \multicolumn{1}{c|}{a16/b8} &           &                        &          &                        &           &              &          &                  
\end{tabular}
\end{table*}}

\subsection{Effect of Quantization on Accuracy}
To see the effect of quantization and the overall accuracy, results with three different quantization levels and various configurations are presented in \figref{final_result}. With a polynomial degree $N \geq 8192$, the encrypted inference shows no degradation in accuracy compared with the unencrypted case. Models with a polynomial degree of $N=4096$ starts to lose accuracy compared to its unencrypted baseline. Though models with a larger hypervector dimension have a better baseline accuracy, more errors are accumulated during the dot-product evaluation, resulting in lower accuracy. The accuracy can be restored by quantizing the class hypervectors to lower precision (e.g., 16 bits). The model with $D=4096$ and $N=4096$ has the best accuracy of 97.10\%. At a polynomial degree of $N=2048$, only a modulus of bit-length 27 is supported, extremely constraining the precision. For every hypervector dimension value, floating-point models have an accuracy under 10\%. Even though we have tried quantizing the class hypervectors to 8 bits, the accuracy still remains below 90\%.

\subsection{Comparison with Other HE-based PPML}

\begin{table}[t!]
  \caption{Comparison with Previous Work}
  \label{comparison}
  \centering
 % \resizebox{\columnwidth}{!}{
  \begin{tabular}{c| c c c c }
  \toprule
  \multirow{2}{*}{Method} & Accuracy & Message &
  \multicolumn{2}{c}{Latency (s)} \\
  \cmidrule{4-5}
  &  (\%) & Size (B) & Encryption & Inference \\
  \midrule
  CryptoNets~\cite{cryptonets:icml16}  &   98.95  &  367.5M$^\dagger$  &    44.5$^\dagger$  &  250$^\dagger$  \\
  LoLa~\cite{brutzkus2019low}         &   98.95  &  11.72M*  &    1.42*  &  2.2 $\ddagger$  \\
  LoLa-Small~\cite{brutzkus2019low}   &   96.92  &  11.72M*  &    1.42*  &  0.29 $\ddagger$ \\
  HE-HDC (Ours)    &   97.10*  &  96.16K*  &    0.011*\footnote[4]{} &  0.083* \\
  \bottomrule
\multicolumn{5}{l}{\emph{Note.}~ 1. $^\dagger$for a batch size of 4096 (other works use the batch size of 1).}\\
\multicolumn{5}{l}{~~~~~~~ 2. $^\ddagger$with multithreading enabled.}\\
\multicolumn{5}{l}{~~~~~~~  3. *Our own measurements.}\\
\multicolumn{5}{l}{~~~~~~~  4. \footnote[4]{}Including the hypervector encoding latency.}
  \end{tabular}
\end{table}

\tabref{comparison} compares our HE-HDC with previous HE-based PPML approaches. The table reports the inference and encryption latency, the message size (the size of ciphertexts sent to the server), and the model accuracy on the MNIST dataset. The encryption latency includes the hypervector encoding latency in the case of HE-HDC. For HE-HDC, we use the configuration of $N=4096$ and $D=8192$, which gives the best accuracy. Our results clearly demonstrate that our HE-HDC outperforms all the previous DNN-based methods by $26\sim3000\times$ in latency with marginal difference in accuracy. This is mainly because HE-HDC uses a polynomial degree of $2^{12}$ and the shortest modulo chain. It is also notable that HE-HDC has the smallest message size and extremely low encryption latency. HE-HDC requires $1\sim 4$ ciphertexts depending on the hypervector dimension whereas LoLa \cite{brutzkus2019low} requires 25 ciphertexts to encrypt a single MNIST image, and CryptoNets \cite{cryptonets:icml16} encrypts every single pixel into a single ciphertext. The latency of single inference on CryptoNets \cite{cryptonets:icml16} is 250 seconds, which needs a batch size of 4096 for full utilization. This may not be realistic, since the entire batch must come from a single client sharing the same secret key.

\ignore{
\begin{table}[t]
  \centering
  \caption{comparison...}
\begin{tabular}{c|ccr@{.}lcl}
\toprule
Method            & Accuracy (\%) & \#Thrds & \multicolumn{2}{c}{Latency (s)} & Note \\ \midrule
CryptoNets        & 98.95         & 1 & 205 &  &         \\ 
FHE–DiNN100       & 96.35         & 1 & 1 & 65 &        \\ 
Faster-CryptoNets & 98.70         & 1 & 39 & 1 &        \\ 
CryptoNets 2.3    & 98.95         & 1 & 24 & 8 &        \\ 
HCNN              & 99.00         & 1 & 14 & 1 &        \\ 
LoLa              & 98.95         & 1 & 2 & 2  &        \\ 
Our Work          &               & 1 &  0 & x &         \\ \bottomrule
\end{tabular}
\todonum[JL]{add another column for note (\# threads, encryption overhead(comp, data))}
\todonum[JW]{do we need this other then Table III?}
\end{table}
}

\section{Conclusion}

We presented a novel PPML scheme, HE-HDC, which uses HDC together with CKKS. Our experimental results based on a full implementation on top of the SEAL library and real measurements of actual CPU runtime clearly demonstrate that our HE-HDC is indeed CKKS-friendly, which has never been reported before. Our optimization methods and cryptography parameter tuning further expedite the encrypted inference procedure, leading HE-HDC to outperform previous state-of-the-art PPML methods by $26\sim3000\times$ times, without significantly sacrificing inference accuracy. 

%The significant performance benefit of HE-HDC is backed by the similarity search technique tailored for the HE-based privacy-preserving MLaaS. The generality of encoding and decoding procedure enables to defer them to the clients without sacrificing security. Learning about the schemes do not give the client advantages in obtaining the model itself. This leaves the computation of the similarity scores the only task for the MLaaS provider. We find that this computation procedure can be further adapted to perform only one level of multiplication. This optimization particularly improves the performance over the specific HE scheme that HE-HDC uses by enabling it to use the performance-optimal set of parameters.

\section*{Acknowledgments}
This research was supported by the MSIT(Ministry of Science and ICT), Korea, under the ITRC(Information Technology Research Center) support program(IITP-2023-2021-0-01817) supervised by the IITP(Institute for Information \& Communications Technology Planning \& Evaluation), and Samsung Electronics Co., Ltd.